\documentclass{article}

\usepackage{graphicx,caption,cite}
\usepackage{amsmath,amssymb,amsfonts}

\def\grg{Gen. Relat. Gravit.}
\def\GRG{Gen. Relat. Gravit.} 

\def\CQG{Classical Quant. Grav.}

\def\JPA{J. Phys. A: Math. Gen.}
\def\JETP{J. Exp. Theor. Phys.}

\def\JMP{J. Math. Phys.}

\def\LRR{Living Rev. Relat.}

\def\PLA{Phys. Lett. A}
\def\prd{Phys. Rev. D}
\def\PR{Phys. Rev.}
\def\PRL{Phys.\ Rev.\ Lett.}
\def\PRSLA{Proc. Roy. Soc. London A}%

\def\RMP{Rev. Mod. Phys.}

\def\ZfP{Z. Phys. }

\graphicspath{{figures/}}

\newcommand{\trho}{\rho}
\newcommand{\tzeta}{\zeta}

\newcommand{\tdifftot}[2]{\tfrac{{\mathrm d}#1}{{\mathrm d}#2}}
\newcommand{\diffpar}[2]{\frac{{\partial}#1}{{\partial }#2}}

\newcommand{\e}[1]{{\mathrm e}^{#1}}
\newcounter{theoremCounter}
\newtheorem{definition}[theoremCounter]{Definition}

\usepackage[affil-it]{authblk}

\begin{document}

\title{Source integrals for multipole moments in static and axially symmetric
spacetimes}
\author{Norman G{\"u}rlebeck}
\affil{ZARM, University of Bremen, Am Fallturm, 28359 Bremen,
Germany\thanks{\texttt{norman.guerlebeck@zarm.uni-bremen.de}}}

\date{}

\maketitle
\begin{abstract}
In this article, we derive source integrals for multipole moments in axially
symmetric and static spacetimes. The multipole moments can be read off
the asymptotics of the metric close to spatial infinity in a
hypersurface, which is orthogonal to the timelike Killing vector.
Whereas for the evaluation of the source integrals the geometry needs to be
known in a compact region of this hypersurface, which encloses all source, i.e. 
matter as well as singularities. The source integrals can be written either as
volume integrals over such a region or in quasi-local form as integrals over the
surface of that region.
\end{abstract}

\section{Introduction}

In general relativity, there were several definitions of multipoles proposed.
Since this theory is non-linear, it is, however, by no means obvious that this
is at all possible. Thus, it is not surprising that in early works multipoles
were only defined in approximations to general relativity that lead to linear
field equations and allow a classical treatment. The most definitions in
this direction and beyond were covered in Thorne's review \cite{Thorne_1980}.

From the 1960s on, new definitions in the full theory of
isolated bodies\footnote{This means that all sources (matter and black holes)
are located in a sphere of finite radius and the spacetime is assumed to be
asymptotically flat. A precise meaning is given in Se. \ref{sec:Geroch_MM}.}
started emerging. These definitions of multipole moments can roughly be divided
into two classes. In the first, the metric (or quantities derived from it) are
expanded at spacelike or null-like infinity. We will call these \emph{asymptotic
definitions} or \emph{asymptotic multipole moments}.
Amongst these are the definitions of Bondi, Metzner, Sachs and van der Burgh (BMSB)
\cite{Bondi_1962,Sachs_1962}, Geroch and Hansen (GH)
\cite{Geroch_1970,Hansen_1974}, Simon and Beig\footnote{This approach reproduces
the GH multipole moments.} (SB) \cite{Simon_1983}, Janis, Newman and Unti (JNU)
\cite{Janis_1965,Newman_1962}, Thorne \cite{Thorne_1980}, the ADM approach
\cite{Arnowitt_1961} and the Komar integrals \cite{Komar_1959}, for reviews see
\cite{Thorne_1980,Quevedo_1990}. There scope of applicability varies greatly.
Whereas the GH multipole moments are defined only in stationary spacetimes the
BMSB, JNU, Thorne and ADM definitions hold in a more general setting. The Komar
expressions for the mass and the angular momentum on the other hand require
stationarity and stationarity and axially symmetry, respectively. Higher order
multipoles are not defined in the Komar approach.
Despite their conceptual differences, G\"ursel showed in \cite{Guersel_1983} the
equivalence of the GH and Thorne's multipole moments in case the requirements of
both definitions are met. Additionally, the mass and the angular momentum in the
GH, Thorne, ADM and Komar approach can be shown to agree.

In the second class fall multipoles that are determined by the metric in a
compact region. Dixon's definition in \cite{Dixon_1973} falls in this class.
These multipoles are given in the form of source integrals. However, it is not
yet known how they are related with the asymptotic multipole moments. A main
application of these multipoles is in the theory of the motion of test bodies
with internal structure. But for test bodies it is obvious that there cannot be any
such relation between Dixon's multipole moments and the asymptotic multipole
moments. Furthermore, Dixon's definition is in general not applicable if
caustics of geodesics appear inside the source, i.e., if the gravitational field
is too strong compared to a characteristic radius of the source. Ashtekar et al.
defined in \cite{Ashtekar_2004} multipole moments of isolated horizons. These
are also source integrals and require only the knowledge of the interior
geometry of the horizon. In \cite{Ashtekar_2004}, it was also shown that the so
defined multipoles of the Kerr black hole deviate from the GH multipoles. This
effect becomes more pronounced the greater the rotation parameter. In both
cases \cite{Dixon_1973,Ashtekar_2004}, it is interesting to find the relation of
these multipole moments to asymptotically defined GH multipole moments and
their physical interpretation. Source integrals might proof useful in this
respect.

Other definitions, in particular for the quasi-local mass and the quasi-local
angular momentum, can be found in \cite{Szabados_2009}. Here we aim at source
integrals for \emph{all} multipole moments in static spacetimes for arbitrary
sources, i.e., we want to express asymptotic multipole moments by surface or
volume integrals, where the surface envelopes or the volume covers all the
sources of the gravitational field. That such source integrals can be found, is
not at all obvious. This is because of the non-linear nature of the Einstein
equations yielding a gravitational field, which acts again as a source.
To overcome the principle difficulties, we will focus here on static and axially
symmetric spacetimes. In this case, the vacuum Einstein equations can be cast in
an essentially linear form. Additionally, they allow the introduction of a
linear system \cite{Maison_1978,Belinskii_1978,Neugebauer_1979}. The latter
might seem superfluous, if the the former holds. However, we
will derive our source integrals relying solely on the existence of the linear
system and the applicability of the inverse scattering technique. This will
allow us in future work to apply the same formalism to stationary and axially
symmetric isolated systems. This is especially relevant for the description of
relativistic stars, where many applications are. A generalization to
spacetimes with an Einstein-Maxwell field exterior to the sources seems also
feasible. In all these cases, multipole moments of the GH type exist.

Source integrals will prove useful in many respects, in particular in the search
of global solutions of Einstein equations describing figures of equilibrium or
relativistic stars, for recent efforts see, e.g.,
\cite{Boshkayev_2012,Bradley_2007,Meinel_2008,Cabezas_2007} and references
therein. For example, an exterior solution can be constructed for a known
interior solution by employing the source integrals to calculate its multipole
moments. From these the exterior solution can be completely determined. This
yields an exterior solution, which does not necessarily match to the interior
solution. However, if it does not, this construction shows that there is
\emph{no} asymptotically flat solution, which can be matched to the given
interior.
Conversely, possible matter sources can be analyzed for given exterior solutions
and their multipole moments. The latter approach is also of astrophysical
interest, since often only the asymptotics of the gravitational field and the
asymptotic multipole moments are accessible to experiments. Source integrals can
be employed to, e.g., restrict the equation of state of a rotating perfect fluid
from observed multipole moments.

Furthermore, source integrals can be used to compare numerical solutions,
analytical solutions and analytical approximations by calculating their
multipole moments\footnote{For difficulties in extracting the multipole moments
from a given numerical metric, see, e.g, \cite{Pappas_2012}.}, see for example
\cite{Manko_2004}. This can also be used to approximate the vacuum exterior of a
given numerical solution by an analytical one, which exhibits the correct
multipole moments up to a prescribed order, see \cite{Teichmueller_2011}. The
merit of source integrals lies in the fact that they determine the multipole
moments using the matter region only, which captures the internal structure of
the relativistic object and is usually determined with high accuracy.
Additionally, source integrals provide the means to test the accuracy of
numerical methods, which are used to determine relativistic stars, cf.
\cite{Stergioulas_2003}, by calculating the multipole moments in two independent
ways: Firstly, using the asymptotics and, secondly, using the source integrals.
This will give also a physical interpretation to possible deviations.

The paper is organized as follows: In Sec. \ref{sec:Preliminaries}, we introduce
the different concepts used later, i.e., the GH and the Weyl multipole moments
as well as the inverse scattering technique. Sec. \ref{sec:source_integrals} is
devoted to the derivation of the source integrals and includes the main results, Eq.
\eqref{eq:source_integrals_Vi}-\eqref{eq:source_integrals_Si}. In Sec.
\ref{sec:Properties}, we will discuss some properties of the obtained source
integrals.

\section{Preliminaries}\label{sec:Preliminaries}

In this section, we will repeat the notions that are needed in the present
paper. Note, that we use geometric units, in which $G=c=1$, where $c$ is the
velocity of light and $G$ Newton's gravitational constant. The metric has the
signature $(-1,1,1,1)$. Greek indices run from $0$ to $3$, lower-case Latin
indices run from $1$ to $3$ and upper-case Latin indices from $1$ to $2$.

\subsection{The line element and the field equation}\label{sec:lineelement}

We consider static and axially symmetric spacetimes admitting a timelike Killing
vector $\xi^a$ and a spacelike Killing vector $\eta^a$, which commutes with
$\xi^a$, has closed timelike curves and vanishes at the axis of rotation. If the
orbits of the so defined isometry group admit orthogonal 2-surfaces, which is
the case in vacuum, in static perfect fluids or static electromagnetic fields,
see, e.g., \cite{Kundt_1966}, then the metric can be written in the Weyl form:
\begin{align}\label{eq:Weyl_line_element}
  ds^2=\mathrm{e}^{2 k-2U}\left(d\rho^2+d\zeta^2\right)+
  W^2\mathrm{e}^{-2U} d\varphi^2-\mathrm{e}^{2U}d t^2,
\end{align}
where the functions $U,~k$ and $W$ depend on $\rho$ and
$\zeta$.\footnote{That this is not the general case, can be seen from the
conformally flat, static and axially symmetric solutions in
\cite{AyonBeato_2006}. The field equations are not imposed there, but they can
be used to define a (rather unphysical) stress energy tensor.} Note that the
metric functions $U$ and $W$ can be expressed by the Killing vectors:
\begin{align}\label{eq:Killing_Vectors}
  \mathrm{e}^{2U}=-\xi_\alpha\xi^\alpha,\quad W^2=-\eta_\alpha\eta^\alpha
  \xi_\beta\xi^\beta.
\end{align}

\newlength{\dummy}
\newlength{\dummyy}
\setlength{\dummy}{\belowdisplayskip}
\setlength{\dummyy}{\abovedisplayskip}

The Einstein equations can be inferred from the ones given in
\cite{Stephani_2003}. Since we will not specify the matter, we give only a
complete set of combinations of the non-vanishing components of the Ricci
tensor:
\begin{subequations}\label{eq:field_equations}
\begin{align}\label{eq:field_equation_general}
  \Delta^{(2)}U+\frac{1}{W}\left(U_{,\trho}W_{,\trho}+
  U_{,\tzeta}W_{,\tzeta}\right)&=\mathrm{e}^{2k-4U}R_{tt},\notag\\
   W_{,\trho\trho}-W_{,\tzeta\tzeta}+2\left(k_{,\tzeta}W_{,\tzeta}-k_{,\trho}W_{,\trho}+
   W\left(U_{,\trho}^2-U_{,\tzeta}^2\right)\right) &=
   W\left(R_{\tzeta\tzeta}-R_{\trho\trho}\right),\notag\\
  W_{,\zeta}k_{,\rho}+W_{,\rho}k_{,\zeta}
  -2WU_{,\zeta} U_{,\rho}-W_{,\trho\tzeta}&=W R_{\trho\tzeta},
\end{align}
\abovedisplayskip0pt 
\begin{align}
  W\Delta^{(2)} W&=\mathrm{e}^{2k-4U}W^2R_{tt}-\mathrm{e}^{2k}
  R_{\varphi\varphi},\notag\\
 -2\Delta^{(2)}k&=\left(R_{\trho\trho}+R_{\tzeta\tzeta}\right)-\mathrm{e}^{2k-4U}
 R_{tt}-\frac{\mathrm{e}^{2k}}{W^2}R_{\varphi\varphi}\notag 
\end{align}
where $\Delta^{(n)}=\left(\tfrac{\partial^2}{\partial\trho^2}+
\tfrac{n-2}{\rho}\tfrac{\partial}{\partial\rho}+
\tfrac{\partial^2}{\partial\zeta^2}\right)$. The fourth equation implies
that we can introduce canonical Weyl coordinates $(\tilde\rho,\tilde\zeta)$ with
$W=\tilde\rho$ via a conformal transformation in vacuum or in all domains, where
$\Delta^{(2)}W=0$ holds, including, e.g., dust. After this coordinate
system is chosen, we drop the tilde again. The remaining coordinate freedom is a
shift of the origin along the symmetry axis, which is characterized by $\rho=0$. Eqs.
\eqref{eq:field_equation_general} simplify in vacuum to the well-known equations
\setlength{\abovedisplayskip}{\dummyy}
\begin{align}\label{eq:field_equation_vacuum}
\begin{split}
  \Delta^{(3)} U&=0,\\
  k_{,\zeta}&=2\rho U_{,\rho}U_{,\zeta},\quad
  k_{,\rho}=\rho\left((U_{,\rho})^2-(U_{,\zeta})^2\right),
\end{split}
\end{align}
\end{subequations}
where $\Delta^{(n)}$ is defined analogously to $\Delta^{(n)}$ but with
canonical Weyl coordinates. The last two equations determine $k$ via a line
integration once $U$ is known. This $k$ automatically satisfies the last
equation in \eqref{eq:field_equation_general}. Hence, only a Laplace equation
for $U$ remains to be solved. Therefore, the Newtonian theory and
general relativity can be treated on the same formal footing. We will do so
here and highlight the difference in Sec. \ref{sec:Properties}. The disadvantage of using the canonical
Weyl coordinates is that they cannot necessarily be introduced in the interior
of the matter, where we have to use other (non-canonical) Weyl coordinates.

\begin{figure}[tbh]
\begin{center}
\includegraphics[scale=0.3]{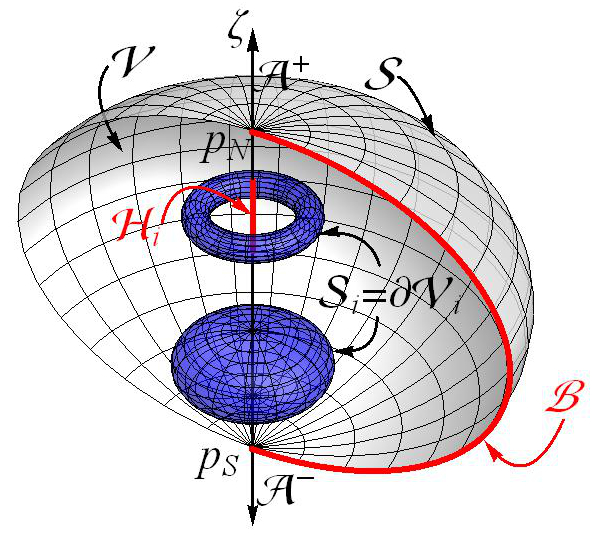}
\caption{The different surfaces $\mathcal S$ and $\mathcal S_i$, volumes $\mathcal V$ and
$\mathcal V_i$ are depicted for a certain matter distributions (ellipsoid and
torus). The curves $\mathcal B$ and $\mathcal A^\pm$ are relevant in Sec.
\ref{sec:linear_problem_Laplace}. Not that this serves only
as illustration and that under a certain energy condition and
separation condition static $n$-body solutions do not exist, for details see \cite{Beig_2009}.
\label{Fig:Curves}}
\end{center}
\end{figure}
The physical setting, which we want to investigate, is as follows: We have in a
compact region $\mathcal V$ of a hypersurface given by $t={\mathrm{const.}}$
several sources, cf. Fig. \ref{Fig:Curves}. The surface of $\mathcal V$ will be
denoted by $\mathcal S=\partial \mathcal V$ and is assumed to be a topological
2-sphere. Note that inside $\mathcal V$ it will in general be not possible to
introduce canonical Weyl coordinates. The sources can be matter distributions
supported in $\mathcal V_i$ with $\mathcal S_i=\partial \mathcal V_i$.
If there are black hole sources with horizons $\mathcal H_i$, we assume that
there is vacuum at least in a neighborhood of $\mathcal H_i$, which necessary
for static black holes, see, e.g., \cite{Bardeen_1973}. Hence, we can define a
closed surfaces $\mathcal S^\mathcal H_i$ in the vacuum neighborhood, which
encloses only the black hole with $\mathcal H_i$.\footnote{Similarly, we can
introduce $\mathcal S_i$ around other singularities, if they have a vacuum
neighborhood.} For relativistic stars, we can choose $\mathcal V=\mathcal V_1$,
in which case we assume that Israel's junction conditions are satisfied across
$\mathcal S=\mathcal S_1$, i.e., that there are no surface distributions. In
fact we will assume this for all surfaces $\mathcal S_i$ and $\mathcal
S_i^\mathcal H$. The only other restriction on the matter is that we want to be
able to introduce Weyl's line element, see \eqref{eq:Weyl_line_element}.

\subsection{Geroch's multipole moments}\label{sec:Geroch_MM}

Isolated gravitating systems are described by an asymptotically flat spacetime.
The precise meaning of this is defined below and used to define Geroch's
multipole moments. Let us denote our static spacetime by $(M,g)$ with the metric
$g_{\alpha\beta}$ and by $V$ a hypersurface orthogonal to $\xi^\alpha$ endowed
with the induced metric
$h_{\alpha\beta}=-\xi^\gamma\xi_\gamma g_{\alpha\beta}+\xi_\alpha\xi_\beta$,
for which we will use subsequently Latin indices.
For the definition of Geroch's multipole moments, asymptotic flatness of $V$ is
sufficient, see \cite{Geroch_1970}:
\begin{definition}
$V$ is asymptotically flat iff there exists a point $\Lambda$, a manifold
$\tilde V$ and a conformal factor $\Omega\in C^2(\tilde V)$, such that
\begin{enumerate}
  \item $\tilde V=V\cup\Lambda$
  \item $\tilde h_{ab}=\Omega^2 h_{ab}$ is a smooth metric of $\tilde V$
  \item $\Omega=\tilde D_a\Omega=0$ and $\tilde D_a\tilde D_b \Omega=2\tilde
  h_{ab}$ at $\Lambda$, where $\tilde D_a$ is the covariant derivative in
  $\tilde V$ associated with the metric $\tilde h_{ab}$.
\end{enumerate}
\end{definition}

Let us define the potential $\tilde \psi$
\begin{align}
 \tilde\psi=\frac{1-(-\xi^\alpha\xi_\alpha)^{\frac12}}{\Omega^{\frac 12}},
\end{align}
which is also a scalar in $\tilde V$. If we introduce the Ricci
tensor $\tilde R^{(3)}_{ab}$ built from the metric $\tilde h_{ab}$
then the tensors $P_{a_1\ldots a_n}$ can be defined recursively:
\begin{align}
\begin{split}
 P&=\tilde \psi\\
 P_{a_1\ldots a_n}&=C\left[\tilde D_{a_1}P_{a_2\ldots
 a_n}-\frac{(n-1)(2n-3)}{2} \tilde R^{(3)}_{a_1a_2}P_{a_3\ldots a_n}\right],
\end{split}
\end{align}
where $C[A_{a_1\ldots a_n}]$ denotes the symmetric and trace-free
part of $A_{a_1\ldots a_n}$. The tensors $P_{a_1\ldots a_n}$ evaluated at
$\Lambda$ define Geroch's multipole moments:
\begin{align}
 M_{a_1\ldots a_n}=\left.P_{a_1\ldots a_n}\right|_{\Lambda}.
\end{align}
The degree of freedom in the choice of the conformal factor reflects the choice
of an origin, with respect to which the multipole moments are taken, see
\cite{Geroch_1970}. 

Up to now, only stationarity was required. If we choose axially symmetry the
multipole structure simplifies to
\begin{align}
 m_r=\left.\frac{1}{r!}P_{a_1\ldots a_r}\tilde z^{a_1}\cdots
 \tilde z^{a_r}\right|_\Lambda,
\end{align} 
where the $\tilde z^a$ is the unit vector pointing in the
direction of the symmetry axis and the scalars $m_n$ define the multipole moments
completely. Hence, we will refer to them as multipole moments, as
well.

Fodor et al. demonstrated in \cite{Fodor_1989} that the multipole moments can be
obtained directly from the Ernst potential on the axis or in the here considered static case
from $U$ at the axis. If we expand $U$ along the axis, i.e.,
$U(\rho=0,\zeta)=\sum\limits^{\infty}_{r=1}U^{(r)}|\zeta|^{-r-1}$, we
characterise the solution by the set of constants $U^{(r)}$. The result in \cite{Fodor_1989}
relates now the $m_r$ with the $U^{(r)}$, i.e., if the $U^{(r)}$ are
known, we can in principle obtain the $m_r$. Thus, we will limit ourselves to
discussing mainly the $U^{(r)}$. Although the relation $m_r(U^{(j)})$ can be
obtained in principal to any order, up to now only the $m_0,\ldots,m_{10}$ were
explicitly expressed using the $U^{(r)}$, see \cite{Fodor_1989}.
We will give here only the first four for illustration:
\begin{align}\label{eq:Fodor_vs_Weyl}
\begin{split}
 m_0=-U^{(0)},\quad m_1=-U^{(1)},\quad m_2=\frac13 {U^{(0)}}^3-U^{(2)},\quad
 m_3={U^{(0)}}^2U^{(1)}-U^{(3)}
\end{split}
\end{align}
Eq. \eqref{eq:Fodor_vs_Weyl} shows that the mass dipole moment,
$U^{(1)}$, can be transformed away, if the mass, $U^{(0)}$, is not vanishing. For a general
discussion, further references and expressions of the center of mass, see
\cite{Cederbaum_2012}.

In \cite{HernandezPastora_2010} a method to obtain the $m_r$ was proposed, which
could help to overcome the non explicit structure of $m_r(U^{(j)})$. Also the
pure $2^r$ pole solutions in \cite{Baeckdahl_2005} could proof useful in
this respect. 

\subsection{The linear problem of the Laplace
equation}\label{sec:linear_problem_Laplace}

Lastly, we shortly review the linear problem associated with the Laplace
equation. Although the equations involved are fairly simple, we decided to use
this technique, because it is readily generalizable to the stationary case.

In the more general case of stationarity and axially symmetry, the Einstein
equation, i.e., the Ernst equation, admits a linear problem, see, e.g.,
\cite{Maison_1978,Belinskii_1978,Neugebauer_1979} and for a recent account
\cite{Neugebauer_2003}. In the static case\footnote{The formulas can easily
inferred from \cite{Neugebauer_2003} by setting $g_{t\varphi}=0$.} the linear problem reads
\begin{align}\label{eq:linear_problem_Laplace}
  \sigma_{,z}=(1+\lambda)U_{,z} \sigma, \quad
  \sigma_{,\bar z}=\left(1+\frac{1}{\lambda}\right)U_{,\bar z} \sigma,
\end{align}
where $z=\rho+{\mathrm i} \zeta$, the spectral parameter
$\lambda=\sqrt{\tfrac{K-\mathrm{i}\bar z}{K+\mathrm{i}z}}$, $K\in \mathbb C$ and
a bar denotes complex conjugation. The function $\sigma$ depends on
$z,~\bar z$ and $\lambda$. The integrability condition of Eqs.
\eqref{eq:linear_problem_Laplace} is the first equation in Eq.
\eqref{eq:field_equation_vacuum}. 

Next we will repeat some known properties of $\sigma$ without proof. For details
we refer the reader to \cite{Neugebauer_2003}. There are four curves of
particular interest $\mathcal A^{\pm}$, $\mathcal B$ and $\mathcal C$, cf. Fig.
\ref{Fig:Curves}. The axis of symmetry is divided by $\mathcal V$ in an upper
and lower part $\mathcal A^+$ and $\mathcal A^-$, respectively. The curve
$\mathcal B$ generates $\mathcal S$ by an rotation around the axis and is given
by a restriction of $\mathcal S$ to $\varphi=0$. Thus, we will subsequently
refer to $\mathcal A^\pm$ and $\mathcal B$ as curves in a $\rho,\zeta$-plane.
Lastly, $\mathcal C$ describes a half circle with sufficiently large radius
connecting $\mathcal A^+$ with $\mathcal A^-$.

Along $\mathcal A^\pm$ and $C$ Eq. \eqref{eq:linear_problem_Laplace} can be
integrated. This yields for a suitable choice of the constant of
integration
\begin{align}\label{eq:axis_values_sigma}
\begin{split}
(0,\zeta)\in \mathcal A^+:~\sigma\left(\lambda=+1,\rho=0,\zeta\right)
&=F(K)\mathrm{e}^{2U(\rho=0,\zeta)},\\
\sigma\left(\lambda=-1,\rho=0,\zeta\right)&=1,\\
(0,\zeta)\in \mathcal A^-:~\sigma\left(\lambda=+1,\rho=0,\zeta\right)
&=\mathrm{e}^{2U(\rho=0,\zeta)},\\
\sigma\left(\lambda=-1,\rho=0,\zeta\right)&=F(K).
\end{split}
\end{align}
The function $F:\mathbb C\to \mathbb C$ is given for $K\in \mathbb
R$ with $(\rho=0,\zeta=K)\in { \mathcal A^{\pm}}$ by
\begin{align}
  F(K)=
  \begin{cases}
\mathrm{e}^{-2U(\rho=0,\zeta=K)} & (0,K)\in {\mathcal A}^+\\
\mathrm{e}^{2U(\rho=0,\zeta=K)} & (0,K)\in {\mathcal A}^-
\end{cases}.
\end{align}
The integration along $\mathcal B$ is the crucial part for our
considerations in the next section.

\section{Source integrals Geroch's multipole
moments}\label{sec:source_integrals}

Let us assume that the line element is written in canonical Weyl coordinates in
the exterior region $\mathcal V^C$, cf. Sec. \ref{sec:lineelement}. The scalars
$U$ and $W$, cf. \eqref{eq:Killing_Vectors}, are also scalars in the projection
$\mathcal V$. Furthermore, $U$ and $W$ are supposed to be continuously
differentiable in the vacuum region including $\mathcal S$ and, thus, $\mathcal
B$. Then we can consider the linear problem \eqref{eq:linear_problem_Laplace}
also along $\mathcal B$ (after a projection to the $\varphi=0$ plane):
\begin{align}\label{eq:Linear_Proble_Surface}
  \sigma_{,s}=\left[U_{,A}s^A+\frac
  12\left(\left(\frac{1}{\lambda}+\lambda\right)U_{,A}s^A+{\mathrm i}
  \left(\frac{1}{\lambda}-\lambda\right)U_{,A}n^A\right)\right]\sigma,
\end{align}
where
$(s^A)=(s^\rho,s^\zeta)=(\tdifftot{\rho}{s},\tdifftot{\zeta}{s})$ and
$(n^A)=(n^\rho,n^\zeta)=(-\tdifftot{\zeta}{s},\tdifftot{\rho}{s})$
denote the tangential vector and the outward pointing normal vector to
the curve $\mathcal B: s\in[s_N,s_S]\to (\rho(s),\zeta(s))$, respectively. The
parameter values $s_{N/S}$ give the ``north/south'' pole
$p_{N/S}$, i.e., $(\rho=0,\zeta=\zeta_{N/S})$, cf.
Fig. \ref{Fig:Curves}. Note that the tangential and the normal vectors are not
necessarily normalized allowing an arbitrary parametrization of $\mathcal B$.

Eq. \eqref{eq:Linear_Proble_Surface} constitutes an ordinary differential
equation of first order with the boundary conditions as given in
\eqref{eq:axis_values_sigma} assuming $(0,K)\in \mathbb \mathcal \mathcal A^{+}\cup
\mathcal A^{-}$. As such it is an overdetermined system, which corresponds to the
integrability of Eqs. \eqref{eq:linear_problem_Laplace}.
However, Eq. \eqref{eq:Linear_Proble_Surface} is readily integrated and the
compatibility condition of the boundary conditions reads
\begin{align}\label{eq:Compatibility_Condition}
\begin{split}
 U(0,K)=&\frac{1}{2}(U(0,\zeta_N)-U(0,\zeta_S))+\\
&\frac{1}{4}\int\limits_{s_N}^{s_S}\left(\left(\lambda^{-1}
+\lambda\right)U_{,A}s^A +\mathrm i\left(\lambda^{-1}-
\lambda\right)U_{,A}n^A\right){\mathrm d} s.
\end{split}
\end{align}
Eq. \eqref{eq:Compatibility_Condition} determines the axis
values of $U$ from the Dirichlet data and the Neumann data along an arbitrary
curve $\mathcal B$, which is sufficient to obtain the entire solution $U$ in
$\mathcal V^C$.

The multipole moments follow from an expansion of Eq.
\eqref{eq:Compatibility_Condition} with respect to $K^{-1}$. Let us denote by
$f^{(r)}$ the expansion coefficient to order $|K|^{-r-1}$ of a function $f(K)$,
which is constant at infinity, i.e.,
$f(K)=\sum\limits_{r=-1}^{\infty}f^{(r)}|K|^{-r-1}$.
The coefficients $N^{(r)}_{+}=(\lambda^{-1}+ \lambda)^{(r)}$ and
$N^{(r)}_{-}={\mathrm i}(\lambda^{-1} - \lambda)^{(r)}$ depend still on
$(\rho,\zeta)$ and satisfy the equations
\begin{align}\label{eq:differential_equations_N}
\begin{split}
 N^{(r)}_{+,\rho}-N^{(r)}_{-,\zeta}&=0,\\
 N^{(r)}_{+,\zeta}+N^{(r)}_{-,\rho}-\frac{1}{\rho}N^{(r)}_{-}&=0. 
\end{split}
\end{align}
Eqs. \eqref{eq:differential_equations_N} follow directly from the form of the
spectral parameter $\lambda$, cf.
after Eq. \eqref{eq:linear_problem_Laplace}. This expansion is only valid for
$\rho^2+\zeta^2<\infty$. Furthermore, $N^{(r)}_{\pm}$ and their radial
derivatives evaluate at the axis to
\begin{align}\label{eq:axis_values_N}
\begin{split}
N^{(r)}_{-}(\rho=0,\zeta)&=0\quad\forall r\geq -1,\\
N^{(-1)}_{+}(\rho=0,\zeta)&=2,\\
N^{(r)}_{+}(\rho=0,\zeta)&=0\quad\forall r\geq 0,\\
N^{(-1)}_{-,\rho}(\rho=0,\zeta)&=0\\
N^{(r)}_{-,\rho}(\rho=0,\zeta)&=-2\zeta^{r}\quad\forall r\geq 0,\\
\end{split}
\end{align} 
The zeroth order of Eq. \eqref{eq:Compatibility_Condition} implies together with
Eq. \eqref{eq:axis_values_N} that $U^{(-1)}=0$. This is also required by
asymptotic flatness, which is assumed in the derivation of
\eqref{eq:axis_values_sigma}. Solving the Eqs.
\eqref{eq:differential_equations_N} yield after a lengthy calculation
$N_{\pm}^{(r)}$ for $r\geq 0$ everywhere:
\begin{align}\label{eq:expansion_coefficients}
\begin{split}
  N_{-}^{(r)}&=
  \sum\limits_{k=0}^{\left\lfloor\frac{r}{2}\right\rfloor}\frac{2(-1)^{k+1}
  r!\rho^{2k+1}\zeta^{r-2k}}{4^k (k!)^2(r-2k)!}\\
  N_{+}^{(r)}&=
  \sum\limits_{k=0}^{\left\lfloor\frac{r-1}{2}\right\rfloor}\frac{2
  (-1)^{k+1}r!\rho^{2k+2}\zeta^{r-2k-1}}{4^k (k!)^2(r-2k-1)!(2k+2)}.
\end{split}
\end{align}
Note that $\tfrac{N^{(r)}_\pm}{\rho}$ is well-behaved also for $\rho\to 0$ for
$r\geq 0$.

For orders $r\geq 0$ Eq. \eqref{eq:Compatibility_Condition} yields together with
Eq. \eqref{eq:expansion_coefficients} the following line integrals defining
Weyl's multipole moments
\begin{align}\label{eq:multipole_moments}
  U^{(r)}=\frac{1}{4}\int\limits_{\gamma_\mathcal B}\left(N^{(r)}_+
  U_{,A}\hat s^A+ N^{(r)}_{-} U_{,A}\hat n^A\right)\mathrm d
  \gamma,
\end{align}
where $\hat s^A$ and $\hat n^A$ are the normalized vectors $s^A$
and $n^A$, respectively, and ${\mathrm d} \gamma$ denotes the proper distance
along the path $\gamma_\mathcal B$, which runs along $\mathcal B$ from the north to the south pole. The functions
$N_{\pm}^{(r)}$ and $U$ are to be read as functions of $(\rho(s),\zeta(s))$.

Eqs. \eqref{eq:multipole_moments} are already expressions of the kind we are
searching for, since they determine the multipole moments from the metric given
in a compact region, i.e., they are quasi-local. But we also can rewrite these
multipole moments as volume integrals justifying the term source integrals even
better. The main obstacle for doing so is that Weyl's multipole moments are
given in Eq. \eqref{eq:multipole_moments} using canonical Weyl coordinates.
Hence, neither the coordinate invariance of these expressions is transparent nor
is obvious how to continue $\rho$ and $\zeta$ to $\mathcal V$. However, $U$ can
be expressed by the norm of the timelike Killing vector, cf. Eq.
\eqref{eq:Killing_Vectors},  which can easily be continued to the interior.

Let us introduce the $1$-form
\begin{align}\label{eq:definition_V}
  Z_\alpha=\epsilon_{\alpha\beta\gamma\delta}W^{,\beta}W^{-1}\eta^\gamma\xi^\delta,
\end{align}
where $\epsilon_{\alpha\beta\gamma\delta}$ denotes the volume form of the static
spacetime.
$Z_\alpha$ is closed in $\mathcal V^C$ and hypersurface orthogonal in the entire
spacetimes. Thus, we can introduce a scalar potential $Z$ with $Z_{,\alpha}=X
Z_\alpha$, where the scalar $X$ equals $1$ in the vacuum region. In canonical
Weyl coordinates in $\mathcal V^{C}$, the potential $Z$ has the trivial form
$Z=\zeta+\zeta_0$. Because we did not fix the origin of our Weyl coordinates,
e.g., the value of $\zeta_N$, we can set the constant $\zeta_0=0$ without loss
of generality. This integration constant is exactly the
freedom we need to change the origin with respect to which the multipole moments
are defined allowing us to change to the center of mass frame. $W=\rho$
and $Z=\zeta$ in $\mathcal V^C$ and they are defined everywhere. Hence, we can
use these two scalars to continue $\rho$ and $\zeta$ into $\mathcal V$. Note
that $W_{,\alpha}$ and $Z_{,\alpha}$ are orthogonal everywhere and have the same
norm in the $\mathcal V^C$. The line integral \eqref{eq:multipole_moments} in
the covariant form reads now
\begin{align}\label{eq:surface_integral_gr}
  U^{(r)}=\frac{1}{4}\int\limits_{\gamma_\mathcal B} N^{(r)}_+(W(s),Z(s)) U_{,a}\hat s^a+ 
  N^{(r)}_{-}(W(s),Z(s)) U_{,a}\hat n^a \mathrm d\gamma.
\end{align}
The dependence on $W(s)$ and $Z(s)$ will be suppressed in the following
expressions.

Along $\mathcal B$ the functions $W$ and $Z$ satisfy
\begin{align}\label{eq:Cauchy_Riemman_surface}
 W_{,s}=Z_{,n},\quad W_{,n}=-Z_{,s},
\end{align}
which is a consequence of the field equation and the choice of canonical
Weyl coordinates. After an partial integration, we can rewrite the line
integral as surface integral using the axial symmetry and Eq.
\eqref{eq:Cauchy_Riemman_surface}, which yields
\begin{align}\label{eq:surface_integrals}
 U^{(r)}=\frac{1}{8\pi}\int\limits_{\mathcal S}
     \frac{\e{U}}{W}\left(N_{-}^{(r)}U_{,\hat n}-N^{(r)}_{+,W}Z_{,\hat
     n}U+N^{(r)}_{+,Z}W_{,\hat n}U\right)\mathrm d \mathcal S.
\end{align}
We denote by $\mathrm d \mathcal S$ the proper surface element of $\mathcal
S,~\mathcal S_i$ or $\mathcal S_i^\mathcal H$, respectively. Since we ruled out
surface distributions, Israel's junction conditions imply that Eqs.
\eqref{eq:surface_integrals} can be understood as integrals over the 2-surface
$\mathcal S$ as seen from the exterior or the interior, see
\cite{Israel_1966}.

Using Stoke's theorem and Eqs. \eqref{eq:field_equations} we obtain (see Fig.
\ref{Fig:Curves} and the end of Sec. \ref{sec:lineelement} for a description of
the $\mathcal S^\mathcal H_i$ and $\mathcal V_i$)
\begin{align}\label{eq:volume_integral}
\begin{split}
 U^{(r)}=&\frac{1}{8\pi}
 \int\limits_{\mathcal
 V}\e{U}\left[-\frac{N^{(r)}_{-}(W,Z)}{W}R_{\alpha\beta}\frac{\xi^\alpha\xi^\beta}{\xi^\gamma\xi_\gamma}
 + N^{(r)}_{+,Z}(W,Z)U
 \left(\frac{W^{,\alpha}}{W\phantom{{}^\alpha}}\right)_{;\alpha}-\right.\\
 &\left.  N^{(r)}_{+,W}(W,Z)U\left( \frac{Z^{,\alpha}}{W\phantom{{}^\alpha}}
 \right)_{;\alpha} +N^{(r)}_{+,WZ}(W,Z)\frac{U}{W}\big(W^{,\alpha}W_{,\alpha}-
 Z^{,\alpha}Z_{,\alpha}\big)\right]\mathrm d\mathcal V+\\
 &\frac{1}{8\pi}\sum\limits_i \int\limits_{\mathcal S^\mathcal H_i}
     \frac{\e{U}}{W}\left(N_{-}^{(r)}U_{,\hat n}-N^{(r)}_{+,W}Z_{,\hat
     n}U+N^{(r)}_{+,Z}W_{,\hat n}U\right)\mathrm d \mathcal S.
\end{split}
\end{align}
Note that the normal vector $\hat n_i^a$ points outward at $\mathcal S^\mathcal H_i$ and
$\mathrm d \mathcal V$ is the proper volume element of $\mathcal V$ or
$\mathcal V_i$, respectively. The covariant derivative with respect to the
metric $g_{\alpha\beta}$ is denoted by a semicolon. The field
equations in the vacuum region \eqref{eq:field_equation_vacuum} imply that the integrand
vanishes there. Hence, we can write Weyl's multipole moments as the
contributions of the individual sources to the total Weyl moment
\begin{align}\label{eq:source_integrals_Vi}
\begin{split}
U^{(r)}=&\frac{1}{8\pi}\sum\limits_{i}
 \int\limits_{\mathcal
 V_i}\e{U}\left[-\frac{N^{(r)}_{-}(W,Z)}{W}\left(T_{\alpha\beta}-\frac{1}{2}
  Tg_{\alpha\beta}\right) \frac{\xi^\alpha\xi^\beta}{\xi^\gamma\xi_\gamma}
  +\right.\\
  &\left. N^{(r)}_{+,Z}(W,Z)U
  \left(\frac{W^{,\alpha}}{W\phantom{{}^\alpha}}\right)_{;\alpha}- 
  N^{(r)}_{+,W}(W,Z)U\left( \frac{Z^{,\alpha}}{W\phantom{{}^\alpha}}
  \right)_{;\alpha} +\right.\\
 &\left.N^{(r)}_{+,WZ}(W,Z)\frac{U}{W}\big(W^{,\alpha}W_{,\alpha}- Z^{,\alpha}Z_{,\alpha}\big)\right]\mathrm d\mathcal V+\\
 &\frac{1}{8\pi}\sum\limits_i \int\limits_{\mathcal S^\mathcal H_i}
     \frac{\e{U}}{W}\left(N_{-}^{(r)}U_{,\hat n}-N^{(r)}_{+,W}Z_{,\hat
     n}U+N^{(r)}_{+,Z}W_{,\hat n}U\right)\mathrm d \mathcal S\\
 =&\sum\limits_{i}U^{(r)}_i+\sum\limits_{i}U^{(r)H}_i.
\end{split}
\end{align}
The integrals in \eqref{eq:source_integrals_Vi} are the source integrals or
quasi-local expressions for the asymptotically defined Weyl moments and, thus,
for the asymptotically defined Geroch-Hansen multipole moments. Note that the
second derivatives of $W$ in Eq. \eqref{eq:source_integrals_Si} can be expressed by the
energy momentum tensor using Eq. \eqref{eq:field_equation_general}. How the
contributions of the black holes are related to the definitions of the multipole
moments of isolated horizons in \cite{Ashtekar_2004} will be investigated in a
future work as well as the relation of the source integrals of $\mathcal V_i$ to
those of Dixon given in, e.g., \cite{Dixon_1973}.

Since the transformation from Weyl's multipole moments to Geroch's is non-linear
except for the mass and the mass dipole, cf. \eqref{eq:Fodor_vs_Weyl}, there is
no linear superposition of the multipole contributions of the individual sources
to the total Geroch multipole moments. Hence a mixing of the contributions of
the individual sources takes place.

Of course, Eq. \eqref{eq:source_integrals_Vi} can again be rewritten as a sum of
surface integrals:
\begin{align}\label{eq:source_integrals_Si}
\begin{split}
U^{(r)}=&\frac{1}{8\pi}\sum\limits_{i}
 \int\limits_{\mathcal S_i} \frac{\e{U}}{W}\left(N_{-}^{(r)}U_{,\hat
 n}-N^{(r)}_{+,W}Z_{,\hat n}U+N^{(r)}_{+,Z}W_{,\hat n}U\right)\mathrm d \mathcal
 S+\\
 &\frac{1}{8\pi}\sum\limits_i \int_{\mathcal S^\mathcal H_i}
     \frac{\e{U}}{W}\left(N_{-}^{(r)}U_{,\hat n}-N^{(r)}_{+,W}Z_{,\hat
     n}U+N^{(r)}_{+,Z}W_{,\hat n}U\right)\mathrm d \mathcal S.
\end{split}
\end{align}
It is obvious that our choice of continuation of $\rho,~\zeta$ into
$\mathcal V$ is not unique and affects greatly the form of Eqs.
\eqref{eq:surface_integrals}-\eqref{eq:source_integrals_Si}, though not the
value. In fact, any $C^1$ extension of the scalars $W,~Z$ from $\mathcal V^C$
could be chosen. Thus, depending on the applications, other choices might be more
appropriate.

\section{Properties of the source integrals}\label{sec:Properties}

In this section, we will discuss the consequences of Eq.
\eqref{eq:multipole_moments} in more detail in the Newtonian and the general
relativistic case. In the Newtonian case, Eq.
\eqref{eq:multipole_moments} comprises the well known multipole definitions as
we will show in the next section.

\subsection{The Newtonian case}

Suppose $U$ is a solution to the Laplace equation in $\mathcal V^C$ (cf.
Fig 1), then it follows by virtue of Green's theorem from the Dirichlet and
Neumann data:
\begin{align}\label{eq:Greens_theorem}
  U(x)=-\frac{1}{4\pi}\int\limits_{\mathcal S}
  \left(G(x,y)\diffpar{U(y)}{y^a}- U(y)\diffpar{G(x,y)}{y}\right)\hat
  n^a\mathrm d \mathcal S_y,
\end{align}
where $G(x,y)$ denotes an arbitrary Green's function, $\mathrm d \mathcal S_y$ a
surface element of the boundary $\mathcal S$ and $x\in \mathcal V^{C}$. The
integration is over $y\in \mathcal S$ and $\hat n^{a}$ is the inward pointing
with respect to $\mathcal V^C$ unit normal to $\mathcal S$ at $y$.

Eq. \eqref{eq:Greens_theorem} is equivalent to Eq. \eqref{eq:surface_integrals}
in case one restricts $x$ to the axis and makes an expansion in $|x|^{-1}$ for
the special choice $G(x,y)=-\tfrac{1}{4\pi|x-y|}$. This yields surface integrals in
the form of \eqref{eq:surface_integrals} with the factor $\e{U}$ set to 1, a
flat space surface element and with $W=\rho$ and $Z=\zeta$. Thus, the multipole
moments still contain all the information as Eq. \eqref{eq:Greens_theorem} with the
difference that the latter is not accessible for stationary, axially symmetric
and isolated sources in general relativity. This is the reason, why we chose the
approach using the linear system.

If Stoke's theorem is applied for these surface integrals, we arrive at
\begin{align}\label{eq:source_integral_Newtonian}
  U^{(r)}=\frac{\mathrm 1}{8\pi}\int\limits_{\mathcal V}\frac{N^{(r)}_{-}}{\rho}
  \Delta^{(3)}U \mathrm d \mathcal V=\frac{ 1}{2}\int\limits_{\mathcal V}\frac{N^{(r)}_{-}}{\rho}
  \mu \mathrm d \mathcal V,
\end{align}
where we made use of the Poisson equation for the Newtonian gravitational
potential, $\Delta^{(3)} U=4\pi\mu$, with a mass density $\mu$. These are, of
course, the usual multipole moments in source integral form up to a sign. This
justifies the term 'source integrals' also for the equivalently obtained expressions
\eqref{eq:source_integrals_Vi} in curved spacetime.
A comparison with the well-known formulas of Newtonian theory shows
that
\begin{align}
 N^{(k)}_{-}=-2 \rho r^k P_k(\cos\theta),\quad \forall r\geq 0
\end{align}
with polar coordinates $(r,\theta)$ defined as usual: $\rho=r\cos\theta,~\zeta=r
\sin\theta$. $P_k$ denote the Legendre polynomials of the first kind. In the
general relativistic case, these $N^{(k)}_\pm$ depend on the extensions of
$\rho,\zeta$ into the interior and thus become polynomials in the scalars
$W$ and $Z$.

\subsection{The general relativistic case}

In this section, we will discuss some of the properties of the quasi-local
volume integral given in Eq. \eqref{eq:volume_integral}. The Einstein equations
are non-linear and contain already the equations of motion (Bianchi identity).
Hence, we could not expect a result like \eqref{eq:source_integral_Newtonian},
which depends only on the mass density. We rather find source integrals
\eqref{eq:source_integrals_Vi} containing terms that are not expressed
explicitly by the matter distribution (all but the first term). However, all of
these terms vanish in vacuum. They also vanish in matter distributions, for
which we can choose $W=\rho$ in $\mathcal V$. In those case the source integrals
have the same form as in the Newtonian case.

Of course, the first multipole moment coincides with the Geroch mass and, hence,
must coincide with the well-known Komar mass. With Eq. \eqref{eq:Fodor_vs_Weyl}
and Eq. \eqref{eq:axis_values_N} we have $N^{(0)}_+=0$ and $N^{(0)}_-=-2 W$
such that only the first term of the integrand in Eq. 
\eqref{eq:volume_integral} remains:
\begin{align*}
 M=\frac{1}{4\pi}\sum\limits_{i}
 \int\limits_{\mathcal V_i}R_{ab}\frac{\xi^a\xi^b}{\sqrt{-\xi^c\xi_c}}\mathrm
 d\mathcal V+ \frac{1}{4\pi}\sum\limits_i \int_{\mathcal S^\mathcal H_i}
 \e{U}U_{,\hat n}\mathrm d \mathcal S
\end{align*}
This is, of course, exactly Komar's integral of the mass in static spacetimes.
The black hole contributions can also be cast in the standard form:
\begin{align}
M_{S^\mathcal H_i}=\frac{1}{4\pi}\int_{\mathcal S^\mathcal H_i}
 \e{U}U_{,\hat n}\mathrm d \mathcal S=\frac{1}{8\pi}\int_{\mathcal S^\mathcal
 H_i}\epsilon_{\alpha\beta\gamma\delta}\xi^{\alpha;\beta}.
\end{align}

Although the main goal of this paper is to present the derivation and definition
of the source integrals, we will give here a short application. We show that
static, axially symmetric and isolated dust configurations do not exist. This is
an old result\footnote{For more general non-existence results for dust,
see also \cite{Caporali_1978,Gurlebeck_2009,Pfister_2010}.}, but can easily be
recovered using source integrals. This demonstrates also how these quasi-local
expressions can be employed. Static and axially symmetric dust configurations
are characterized by the energy-momentum tensor in Weyl coordinates
\begin{align}
T_{ab}=\mu\e{2U}\delta_a^t\delta_b^t.
\end{align}
The Bianchi identity implies $U_{,a}=0$ in $\mathcal V$ and, thus, at
$\mathcal S$.
This yields together with the quasi-local surface integrals for the Weyl moments Eq.
\eqref{eq:surface_integral_gr}
\begin{align}
  U^{(r)}=0.  
\end{align}
Thus, the system has no mass or any other multipole moment, which implies flat
space in the vacuum region. This is clearly a contradiction to a dust source
with positive mass density.

\section{Conclusions}

We have derived in this article source integrals or quasi-local expressions for
Weyl's multipole moments and, thus, for Geroch's multipole moments for axially
symmetric and stationary sources. These source integrals can either be written
as surface integrals or volume integrals. A priori, one could not expect to find
any kind of source integrals at all, because of the non-linear nature of the
Einstein equations. That this is possible in the here considered setting, seems
not to be due to the staticity and axially symmetry and the peculiarly simple
form of the field equations. But rather a linear system must be available
offering a notion of integrability of the Einstein equation. Thus,
it appears feasible to find source integrals not only for stationary and axially
symmetric isolated systems, which describe vacuum, but also electrovacuum close to spatial
infinity. These generalizations will be investigated in future work.

It should also be clarified, how the source integrals are connected to the
already known source integrals for isolated horizons \cite{Ashtekar_2004}. In
\cite{Ashtekar_2004} it was shown that the source integrals characterize the
horizon uniquely. However, they do not reproduce the GH multipole moments of a
Kerr black hole. In our approach, the agreement of the source integrals and the
asymptotically defined Weyl or Geroch multipole moments is given by
construction. Therefore, these source integrals might prove useful for
identifying the contributions to the multipole moments, which yield the
discrepancies between the isolated horizon multipole moments and those of
Geroch and Hansen.

\section*{Acknowledgment}
N.G. gratefully acknowledges support from the DFG within the Research Training
Group 1620 ``Models of Gravity''. The author thanks C. L\"ammerzahl,
 V. Perlick and O. Sv\'itek for helpful discussions.


\end{document}